\documentclass[%
reprint,
superscriptaddress,
amsmath,amssymb,
aps,
]{revtex4-1}

\usepackage{graphicx}
\usepackage{dcolumn}
\usepackage{bm}
\usepackage{soul}
\usepackage[centerlast]{caption}
\usepackage{float}
\usepackage{tabularx}
\usepackage{array}
\usepackage{multirow}
\newcolumntype{Y}{>{\centering\arraybackslash}X}
\usepackage{siunitx}
\usepackage{hyperref}
\hypersetup{colorlinks=true, citecolor=blue, urlcolor=blue, linkcolor=blue}
\usepackage{todonotes}
\usepackage{xcolor}
\usepackage{colortbl}
\makeatletter
\usepackage{braket}
\def\CT@@do@color{%
	\global\let\CT@do@color\relax
	\@tempdima\wd\z@
	\advance\@tempdima\@tempdimb
	\advance\@tempdima\@tempdimc
	\advance\@tempdimb\tabcolsep
	\advance\@tempdimc\tabcolsep
	\advance\@tempdima2\tabcolsep
	\kern-\@tempdimb
	\leaders\vrule
	\hskip\@tempdima\@plus  1fill
	\kern-\@tempdimc
	\hskip-\wd\z@ \@plus -1fill }
\makeatother
\usepackage{colortbl}

\usepackage{soul}
\usepackage{upgreek}

\begin{document}
	
\title[High--kinetic inductance NbN films for high--quality \\ compact superconducting resonators]%
{High--kinetic inductance NbN films for high--quality \\ compact superconducting resonators}

\author{S.~Frasca}
\email[E-mail: ]{simone.frasca@epfl.ch}
\affiliation{Advanced Quantum Architecture Laboratory (AQUA), \'{E}cole Polytechnique F\'{e}d\'{e}rale de Lausanne (EPFL) at Microcity, 2002 Neuch\^{a}tel, Switzerland.}
\affiliation{Hybrid Quantum Circuit Laboratory (HQC), \\ \'{E}cole Polytechnique F\'{e}d\'{e}rale de Lausanne (EPFL), 1015 Lausanne, Switzerland.}		
\affiliation{Center for Quantum Science and Engineering, \\ \'{E}cole Polytechnique F\'{e}d\'{e}rale de Lausanne (EPFL), 1015 Lausanne, Switzerland}

\author{I.~N.~Arabadzhiev}
\affiliation{Advanced Quantum Architecture Laboratory (AQUA), \'{E}cole Polytechnique F\'{e}d\'{e}rale de Lausanne (EPFL) at Microcity, 2002 Neuch\^{a}tel, Switzerland.}

\author{S.~Y.~Bros~de~Puechredon}
\affiliation{Advanced Quantum Architecture Laboratory (AQUA), \'{E}cole Polytechnique F\'{e}d\'{e}rale de Lausanne (EPFL) at Microcity, 2002 Neuch\^{a}tel, Switzerland.}

\author{F.~Oppliger}
\affiliation{Hybrid Quantum Circuit Laboratory (HQC), \\ \'{E}cole Polytechnique F\'{e}d\'{e}rale de Lausanne (EPFL), 1015 Lausanne, Switzerland.}	
\affiliation{Center for Quantum Science and Engineering, \\ \'{E}cole Polytechnique F\'{e}d\'{e}rale de Lausanne (EPFL), 1015 Lausanne, Switzerland}

\author{V.~Jouanny}
\affiliation{Hybrid Quantum Circuit Laboratory (HQC), \\ \'{E}cole Polytechnique F\'{e}d\'{e}rale de Lausanne (EPFL), 1015 Lausanne, Switzerland.}	
\affiliation{Center for Quantum Science and Engineering, \\ \'{E}cole Polytechnique F\'{e}d\'{e}rale de Lausanne (EPFL), 1015 Lausanne, Switzerland}

\author{R.~Musio}
\affiliation{Advanced Quantum Architecture Laboratory (AQUA), \'{E}cole Polytechnique F\'{e}d\'{e}rale de Lausanne (EPFL) at Microcity, 2002 Neuch\^{a}tel, Switzerland.}

\author{M.~Scigliuzzo}
\affiliation{Laboratory of Photonics and Quantum Measurements (LPQM), \'{E}cole Polytechnique F\'{e}d\'{e}rale de Lausanne (EPFL), 1015 Lausanne, Switzerland.}
\affiliation{Center for Quantum Science and Engineering, \\ \'{E}cole Polytechnique F\'{e}d\'{e}rale de Lausanne (EPFL), 1015 Lausanne, Switzerland}

\author{F.~Minganti}
\affiliation{Laboratory of Theoretical Physics of Nanosystems (LTPN), \'{E}cole Polytechnique F\'{e}d\'{e}rale de Lausanne (EPFL), 1015 Lausanne, Switzerland.}
\affiliation{Center for Quantum Science and Engineering, \\ \'{E}cole Polytechnique F\'{e}d\'{e}rale de Lausanne (EPFL), 1015 Lausanne, Switzerland}

\author{P.~Scarlino}
\affiliation{Hybrid Quantum Circuit Laboratory (HQC), \\ \'{E}cole Polytechnique F\'{e}d\'{e}rale de Lausanne (EPFL), 1015 Lausanne, Switzerland.}		
\affiliation{Center for Quantum Science and Engineering, \\ \'{E}cole Polytechnique F\'{e}d\'{e}rale de Lausanne (EPFL), 1015 Lausanne, Switzerland}

\author{E.~Charbon}
\affiliation{Advanced Quantum Architecture Laboratory (AQUA), \'{E}cole Polytechnique F\'{e}d\'{e}rale de Lausanne (EPFL) at Microcity, 2002 Neuch\^{a}tel, Switzerland.}

\date{\today}

\begin{abstract}

Niobium nitride (NbN) is a particularly promising material for quantum technology applications, as entails the degree of reproducibility necessary for large-scale of superconducting circuits.
We demonstrate that resonators based on NbN thin films present a one-photon internal quality factor above $10^5$ maintaining a high impedance (larger than $2\mathrm{k}\Omega$), with a footprint of approximately $50\times 100 \, \mu \rm{m}^2$ and a self-Kerr nonlinearity of few tenths of Hz.
These quality factors, mostly limited by losses induced by the coupling to two-level systems, have been maintained for kinetic inductances ranging from tenths to hundreds of pH/$\square$. We also demonstrate minimal variations in the performance of the resonators during multiple cooldowns over more than nine months.
Our work proves the versatility of niobium nitride high--kinetic inductance resonators, opening perspectives towards the fabrication of compact, high--impedance and high--quality multimode circuits, with sizable interactions.
\end{abstract}

\maketitle

\section{\label{sec:intro}Introduction}

The possibility of tuning the magnitude of the inductance in superconducting circuits is paramount to achieve the high-degree of control and flexibility required for quantum technology \cite{kjaergaard_2020}.
Two platforms have been mostly used to achieve high values of inductance: arrays of Josephson junctions (JJs) \cite{scigliuzzo_2022,manucharyan_2009,Kuzmin_2019,Roch_2019} or high--kinetic inductance disordered thin films \cite{annunziata_tunable_2010, maleeva_2018}.
JJ devices are characterized by very low dissipation rates, and for this reason they are routinely used in circuit quantum electrodynamics (cQED) \cite{kono_2020,wang_2022}. Usually, JJ-based circuits still exhibit large nonlinearities. To dilute such nonlinearity suited for applications such as quantum--limited parametric amplifiers, arrays of JJs sacrifice compactness and add fabrication overhead \cite{white_2015, macklin_2015}. 

Kinetic inductance is a promising novel resource for quantum technology \cite{winkel_2020,day_mkids_2003,macklin_2015,day_paramp_2012}. For instance, superconducting high--kinetic inductance (high--$L_\text{k}$) thin films have been used for several cryogenic applications, ranging from detectors \cite{day_mkids_2003,goltsman_picosecond_2001,cabrera_tes_1998} to amplifiers \cite{day_paramp_2012, parker_2022}, and filters \cite{liu_dbr_2017,Sigillito_dbr_2017}.
In dirty superconductor thin films, the inductance depends on the Cooper pairs carrier density \cite{annunziata_tunable_2010}, and it can be tuned by controlling the chemical composition of the films during the deposition. 
From a cQED perspective, the advantages of high--$L_\text{k}$ are manifold.
Thin films give the designers the freedom to operate with low-- or high--impedance, and thus a higher range of achievable capacitive-coupling between different circuital elements \cite{Devoret_2007}.
In turn, this allows exploring several regimes of light-matter interaction, from the strong to more exotic ultrastrong couplings \cite{FriskKockum2019, Forn-Diaz2019}.
Given the compact nature of high--$L_\text{k}$ film resonators, thin films can reduce the dimensions of both readout and control apparatus, and facilitate the realization of many-modes quantum devices \cite{niepce_2019}.
Key to the success of this technology within a quantum framework is attaining regimes of low internal losses, where long coherence times can be maintained. 

Many disordered superconductors have shown high--kinetic inductance, e.g., NbN \cite{Anferov_2020,niepce_2019}, NbTiN \cite{parker_2022,samkharadze_2016,day_paramp_2012}, and TiN \cite{leduc_2010}, or materials composed by microscopic effective Josephson arrays such as the emergent granular Aluminum (grAl) \cite{Pop_2019}.
Several groups manufactured weakly-disordered thin films that present considerable quality factors \cite{vissers_2010, leduc_2010}.
Consequently, thin films with moderate $L_\text{k}$ proved useful for quantum computing tasks \cite{oliver_superguide_2019}, in particular for optical communication \cite{Shaw_2017}, quantum key distribution \cite{takesue_2007,Boaron_2018}, and quantum teleportation \cite{Bussieres_2014}. 

\begin{figure*}	
    \centering
    \includegraphics[width=.96\linewidth]{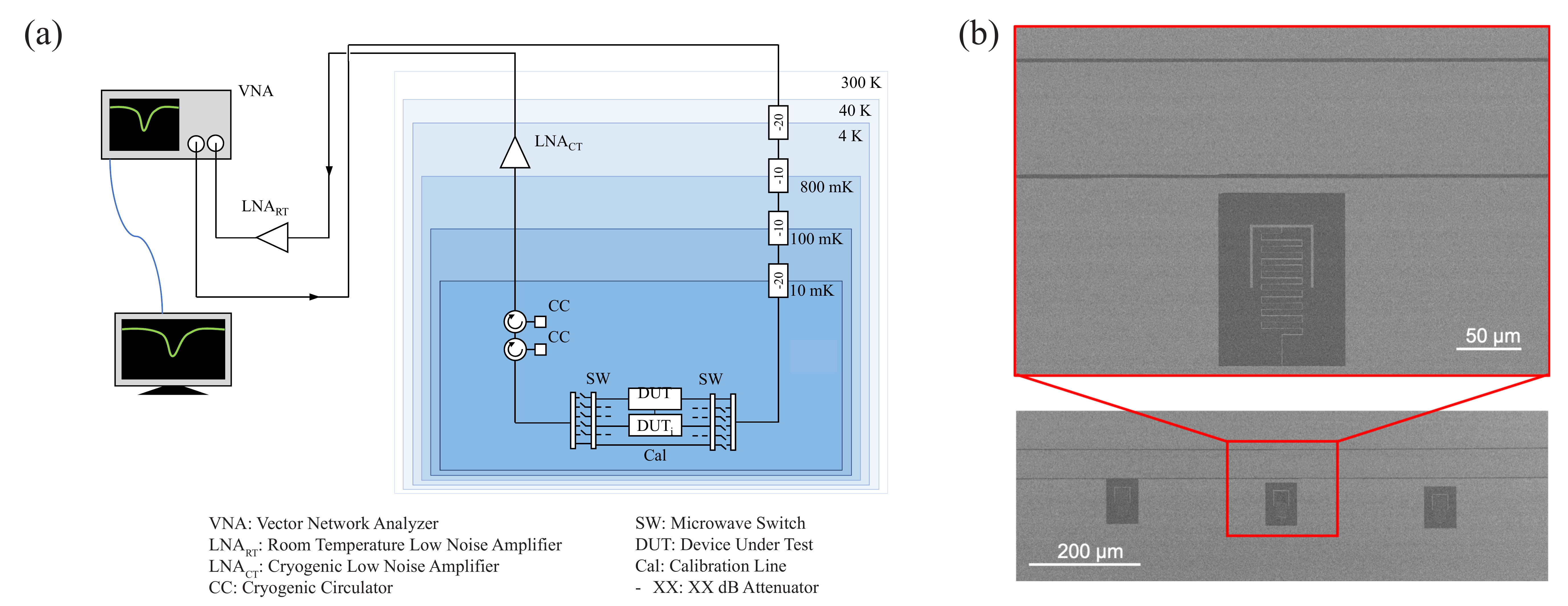}
	\captionsetup{justification=raggedright, singlelinecheck=false}
	\caption{\textbf{(a)} Schematics for the cryogenic setup. The input lines are attenuated with 60~dB cryogenic attenuators distributed across the several stages of the cryostat as in figure.  \textbf{(b)} Scanning electron micrograph of a representative resonator. The hanger resonators are capacitively coupled to a 50~$\Omega$ microwave feedline, also realized in NbN. Each chip consists of two feedlines, each one coupled to seven frequency multiplexed resonators, whose inductor width is either 250~nm or 500~nm.}
	\label{fig:setup}
\end{figure*}

Large-scale quantum applications will, however, also require a wide range of $L_\text{k}$ tunability and high control and reproducibility of the superconducting building blocks. 
In this work, we demonstrate that we fully control the kinetic inductance of NbN films (from $30$ to $170\,\rm{pH}/\square$), maintaining a very-high resonator quality  in excess of $10^5$ at low photon number, with a high and reliable degree of reproducibility.
We identify the saturable two-level systems (TLS) as the main limiting factor to single-photon lifetime \cite{scigliuzzo_2020}. Furthermore, our devices show no significant film degradation (ageing) over the course of different cooldowns nine months apart.

\section{\label{sec:fab}Device Design and Fabrication}

We fabricate planar lumped LC resonators etching 13~nm-thick NbN film, with typical impedance ${Z = \sqrt{L_r/C_r}}$ of 2~k$\Omega$, where $L_r$ and $C_r$ are inductance and capacitance of the resonator respectively. The fabrication begins with a 2~minutes dip in 40\% HF bath to remove the native oxide and possible contamination from the surface of an intrinsic, high--resistivity ($\ge 10~k\Omega$cm), $\braket{100}$--oriented 100~mm Si wafers. There follows an NbN films bias sputtering \cite{dane_bias_2017} at room temperature in a Kenosistec RF sputtering system. After subsequent deposition of Ti/Pt alignment markers by optical lift-off process and dehydration step at 150$^\circ$C for 5~minutes, 80~nm-thick CSAR positive e-beam resist is spin coated at 4000~rpm on the wafer, and baked at 150$^\circ$C for 5~minutes. With an electron beam lithography (Raith EBPG5000+ at 100~keV) step, the devices are patterned on the resist through development in amyl acetate for 1~minute, followed by rinsing in a solution 9:1~MiBK:IPA. The pattern is then transferred to the NbN using CF$_\text{4}$/Ar mixture and reactive ion etching with a power of 15~W for 5 minutes. The resist is stripped by means of Microposit remover~1165 heated at 70$^\circ$C. Finally, the wafer is coated with 1.5~$\upmu$m AZ~ECI~3007 positive photolithography resist for devices protection and diced.

The advantages of bias sputtering \cite{dane_bias_2017}, \textit{i.e.}, application of a RF bias voltage on the substrate, resides in ion bombardment during the film deposition, which causes a reduction of superconducting critical temperature and grain size, permitting the deposition of a polycrystalline material. 
The polycrystallinity improves the superconductor homogeneity and it eases the device realization.
The film behaves as an amorphous material with respect to etching procedures, and simultaneously maintains the advantageous properties of crystalline superconductors.
For example, the low electron--phonon interaction time is barely changed, with important repercussions, such as enhanced maximum count rate in superconducting single--photon detectors \cite{dane_bias_2017}. 
To optimize the films properties, we fine-tune the nitrogen to argon partial pressures in the chamber across several fabrication runs, with pressure and substrate bias chosen to give the lest roughness in the films.
These steps also improve the reproducibility and yield of the samples. 

\section{\label{sec:setup}Experimental Setup}

After the fabrication, we glue the chips on a copper support with PMMA and wire-bond it to a customized printed circuit board. The copper support is then mounted on the cold finger installed at the mixing chamber stage of a BlueFors LD250 dilution refrigerator at a base temperature of 10~mK, see Fig.~\ref{fig:setup}(a). 

In order to test multiple devices in a single cooldown and to increase the experimental throughput, the devices are connected in transmission configuration through cryogenically operating coaxial switches (Radiall, R573 series),  sharing both input and output lines [see Fig.~\ref{fig:setup}(a)].

Each sample is composed by 7 resonators, capacively coupled to a common $50\,\Omega$ coplanar transmission line, as Fig.~\ref{fig:setup} (b). 
The output line is connected to two cryogenic insulators (Low Noise Factory, LNF-ISISC4-8A series) to attenuate thermal noise injection from the amplifier, and a cryogenic low--noise amplifier (Low Noise Factory, LNF-LNC4-8C series) operating at the 4~K stage of the cryostat. 
Input and output lines are then connected at room temperature to a Vector Network Analyzer (Rohde \& Schwarz, ZNA26 series) to acquire the scattering parameters. 

The notch (or hanger \cite{McRae_2020}) configuration of resonators has been chosen for a precise estimation of the internal quality factor of the devices -- they self-calibrate with respect to the transmission baseline -- and it allows frequency multiplexing \cite{McRae_2020}. Moreover, the microwave feedline can be probed in DC to estimate the resistance--temperature characteristic curve and the critical temperature $T_C$. The latter has been measured with a closed loop cryostat (PhotonSpot inc.) with 800~mK base temperature.

\begin{figure}
    \centering
    \captionsetup{justification=raggedright, singlelinecheck=false}
    \includegraphics[width=\linewidth]{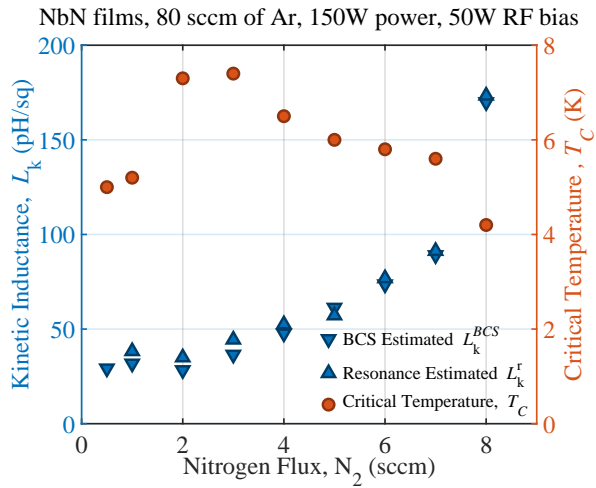}
    \caption{Critical temperature and estimated kinetic inductance for different NbN recipes. All the films presented are 13.0~$\pm$0.6~nm thick. With increasing N$_2$ flux, the films show first a sudden increase of $T_C$ in correspondence of stochiometric conditions, followed by a decrease caused by the large amount of impurities of the dirty superconductor. At the same time, the kinetic inductance first slightly drops due to the increase of $T_C$ as expected from Eq~\eqref{eq:lk}, and then rapidly increases due to both $R_\square$ and $T_C$ contributions. The kinetic inductance was estimated both by BCS theory ($L_\text{k}^{BCS}$) and by fitting the resonant frequency of the resonators with Sonnet ($L_\text{k}^r$) [see Appendix~\ref{app:lk}].}
    \label{fig:films}
\end{figure}

\section{\label{sec:results}Results}
\subsection{NbN films composition, critical temperature and kinetic inductance}

The kinetic inductance per square  $L_\text{k}$  of the film can be expressed as \cite{annunziata_tunable_2010}:
\begin{equation}
    L_\text{k}(T) = \frac{R_\square \hbar}{\pi \Delta}\frac{1}{\tanh{(\frac{\Delta}{2 k_B T})}}.
    \label{eq:lk}
\end{equation}
Here, $R_\square$ is the sheet resistance when the film is in the normal state, $T$ is the superconducting film temperature, and $\Delta$ is the superconducting band gap that, according to BCS theory, can be approximated to $\Delta \simeq \text{1.764}~k_BT_C$ for $T\ll T_C$ with $T_C$ the critical temperature. 

\begin{table}
    \centering
	\captionsetup{justification=raggedright, singlelinecheck=false}
	\caption{Measured and estimated main parameters of the deposited superconducting thin films. The kinetic inductivity of the films is estimated according to the procedure presented in Appendix~\ref{app:lk}.}
	\begin{tabularx}{\linewidth}{XYYYY}
		\hline
	    \multicolumn{5}{>{\hsize=\dimexpr5\hsize+5\tabcolsep+\arrayrulewidth\relax}Y}{\textbf{Film Properties}} \\
	   Ar/N$_{\text{2}}$ & $T_C$ & $R_\square$ & $L_\text{k,0}$ & $j_c$ \\ [-3pt]
		(sccm) & (K) & ($\Omega$/$\square$) & (pH/$\square$) & (A/cm$^\text{2}$) \\
		[.5\normalbaselineskip] \hline
		80/0.5 & 5.0 & 106 & 34.5 & -- \\
		\rowcolor{gray!10}80/1 & 5.2 & 122 & 38.2 & $1.23\times10^{6}$ \\
		80/2 & 7.3 & 151 & 34.9 & $1.36\times10^{6}$ \\
		\rowcolor{gray!10}80/3 & 7.5 & 196 & 44.4 & $1.66\times10^{6}$ \\
		80/4 & 6.5 & 225 & 52.5 & $1.51\times10^{6}$ \\
		\rowcolor{gray!10}80/5 & 6.0 & 267 & 57.2 & $1.21\times10^{6}$ \\
		80/6 & 5.8 & 310 & 76.8 & $1.05\times10^{6}$ \\
		\rowcolor{gray!10}80/7 & 5.6 & 362 & 91.3 & -- \\
		80/8 & 4.2 & 518 & 173.3 & -- \\\hline
	\end{tabularx}
		\label{tab:films}
\end{table}

\begin{figure*}
    \captionsetup{justification=raggedright, singlelinecheck=false}
    \includegraphics[width=.95\linewidth]{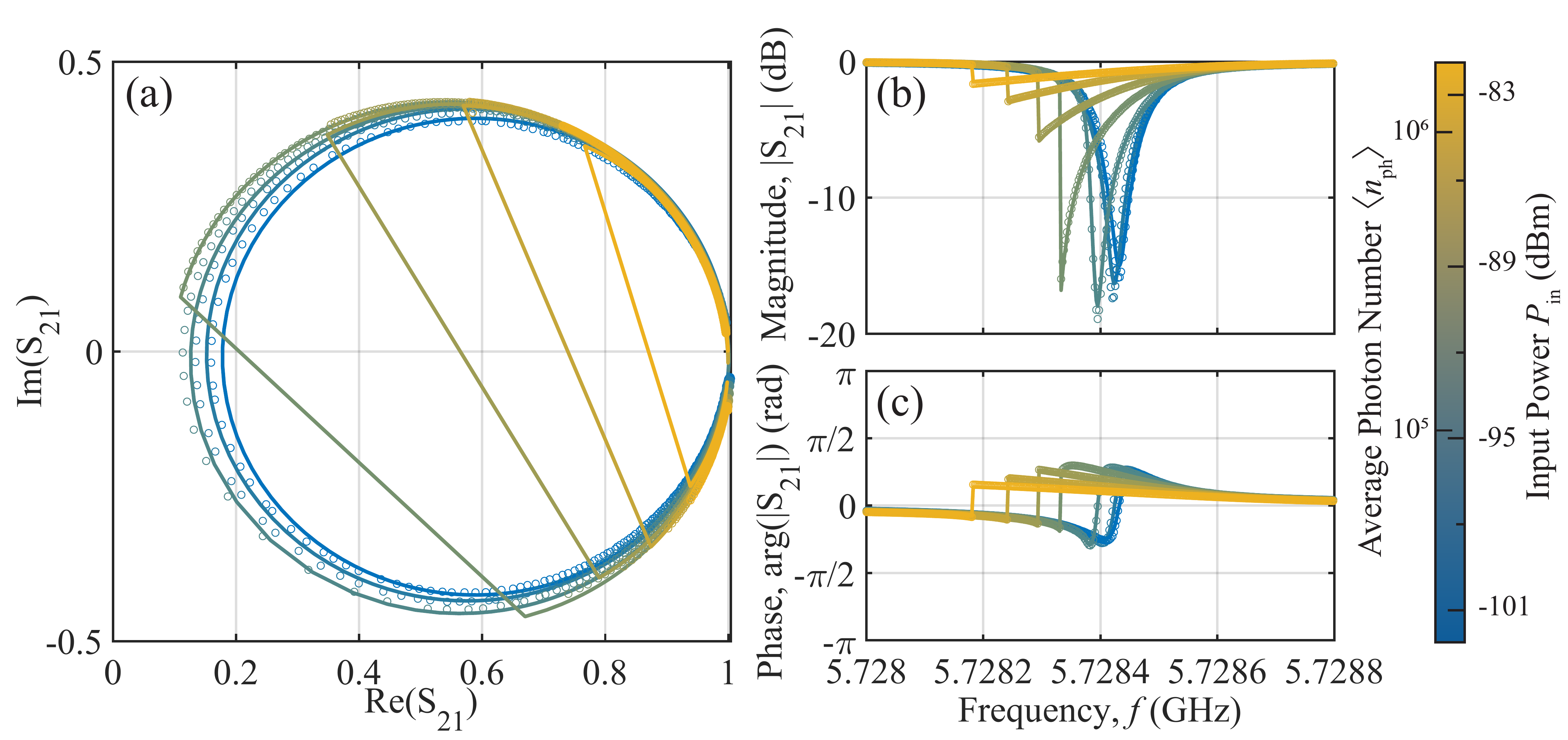}
    \caption{Scattering parameter $S_{21}$ for a representative resonator. \textbf{(a)} Complex plane representation of $S_{21}$ for seven different input powers. \textbf{(b)} Magnitude and \textbf{(c)} phase response as a function of the probe frequency of the same dataset presented in panel \textbf{(a)}. Circles represent datapoints, while solid lines are the fitted functions according to Eq.~\eqref{eq:s21_kerr_hanged}.}
    \label{fig:resonator}
\end{figure*}

Fig.~\ref{fig:films} reports $T_C$ and $L_\text{k}(T)$, both measured and estimated according to Eq.~\eqref{eq:lk}, as a function of the N$_2$ flow during the spattering deposition.
The deposition conditions, critical temperatures, sheet resistance, estimated kinetic inductance, and critical current density $j_c$ of the thin films are reported in Tab.~\ref{tab:films}.
Increasing the N$_2$ concentration in the films causes an increment of impurities embedded in the films during the sputtering process. 
At low N$_2$, we notice a raise in the critical temperature.
While also the resistivity increases, the ratio between these two in Eq.~\eqref{eq:lk} does not compensate, 
causing a little drop in $L_{\rm k}$ in the proximity of the stochiometric condition, which for this deposition setting are found approximately at 2.5~sccm N$_2$ flux. Larger N$_2$ concentration increases the film disorder, causing a raise of film resistivity and a drop of critical temperature, both boosting the kinetic inductance.

\subsection{Internal quality factor and Kerr nonlinearity}

We model each cavity as a Kerr nonlinear resonator, 
described by the Hamiltonian:
\begin{equation}
    H = \hbar \omega_0 a^\dagger a + \hbar \frac{K}{2} (a^\dagger a)^2, \label{eq:Kerr_hamilt}
\end{equation}
where $\hat{a}$ ($\hat{a}^\dagger$) is the bosonic creation (annihilation) operator, $\omega_0$ is the resonant frequency of the cavity, and $K$ the Kerr nonlinearity.
These devices are also characterized by photon loss events, whose rate is given by $(\kappa + \gamma)$, with $\kappa$ the external coupling and $\gamma$ the internal losses.
The resonator is driven at a frequency $\omega_\text{d}$ of intensity $P_{\rm in}$.

For our devices, we measure the transmitted power $S_{21}$ of the resonator hanged to the feedline.
To extract the parameters of a Kerr resonator in hanger configuration, we use input--output theory \cite{Eichler_2014}, and obtain:
\begin{equation}
S_{21} = 1 - \frac{\kappa}{\kappa+\gamma}\frac{e^{i\phi}}{\cos{\phi}}\frac{1}{1+2i(\delta-\xi n)}, \label{eq:s21_kerr_hanged}
\end{equation}
where the interdependent variables $\phi$, $\delta$, and $\xi$ depend on $K$, $\kappa$, $\gamma$, $\omega_0$, on the drive frequency $\omega_\text{d}$, on the photon number $n$, and on other internal parameters of the resonator that can be independently measured, as detailed in Appendix~\ref{app:eq}.
As it will be detailed later, the internal dissipation rate $\gamma$ has a nontrivial power dependence. 
Hence, we perform power scans of the devices to be able to estimate the total loss rate $(\kappa+\gamma)$ with respect to the average number of photons in the resonator $\braket{n_\text{ph}}$. 
We first estimate the resonator frequency $\omega_0$, as it corresponds to the dip in frequency of $S_{21}$ in the low-power regime.
Then, to retrieve $\kappa$, $\gamma$ and $K$, we use a global fit routine at all powers, to increase the accuracy of the extrapolated values.
In the linear regime at low-power, we also benchmark our results for $\kappa$ and $\gamma$ using the python package by Probst \textit{et al.} \cite{Probst_2015}, finding no significant discrepancies with respect to our routine.

In order to precisely estimate the internal quality factor and reduce fit uncertainty, it is more convenient to approach a critical coupling condition $\kappa \approx \gamma$, where the internal quality factor approximately matches the external coupling \cite{McRae_2020}. 
As the internal quality factor is not known a priori, we take full advantage of notch configuration to engineer multiple couplings $\kappa$ for devices on the same chip.
Fourteen hanger resonators are fabricated per each film, seven with inductor width of 250~nm and seven with inductor width of 500~nm. 
In Fig.~\ref{fig:resonator} we present the measured and fitted transmitted power $S_{21}$ using Eq.~\eqref{eq:s21_kerr_hanged} through the feedline for one of the critically coupled resonators as a function of the probe frequency and for several input powers.

\begin{figure*}
    \centering
    \captionsetup{justification=raggedright, singlelinecheck=false}
    \includegraphics[width=.96\linewidth]{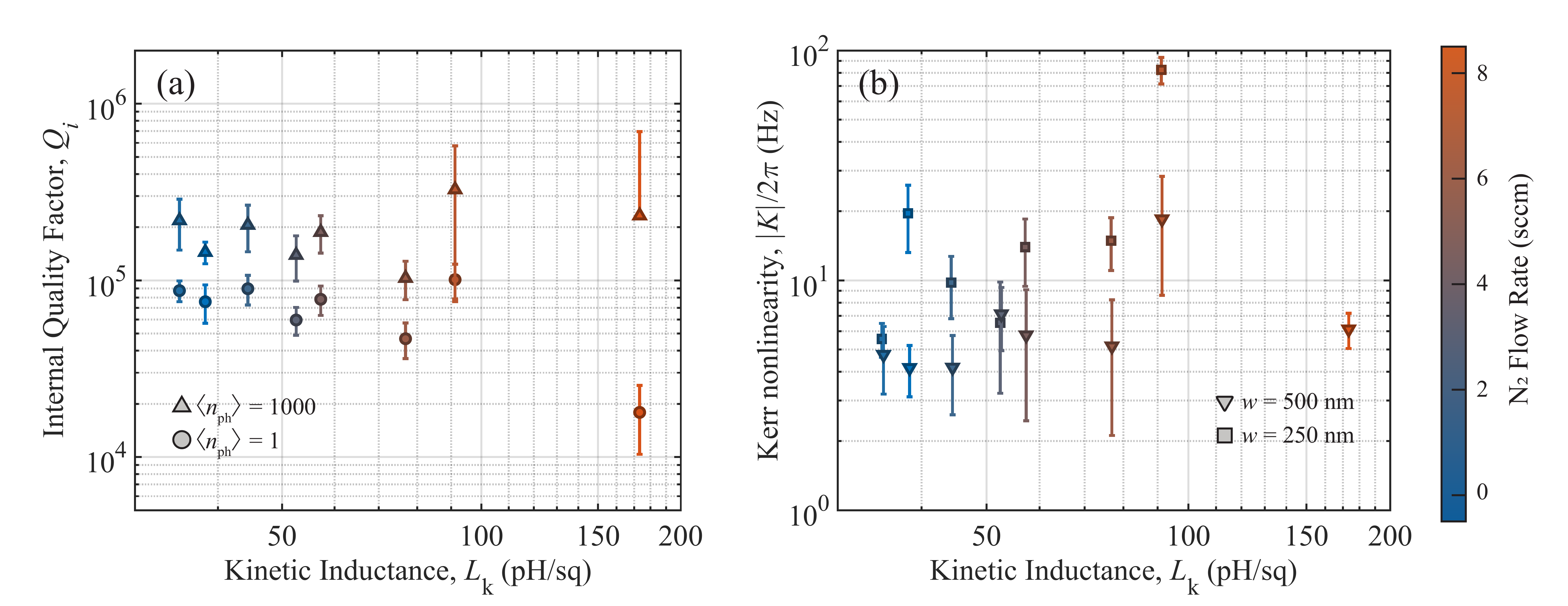}
    \caption{\textbf{(a)} Dependence of the averaged resonators internal quality factor $Q_i$ with respect to kinetic inductance $L_\text{k}$. The extracted $Q_i$ at low photon number ($\braket{n_\text{ph}} \approx 1$) [high photon number ($\braket{n_\text{ph}} \approx 1000$)] are represented with circle [triangle] markers. \textbf{(b)} Extracted self-Kerr nonlinearity $K$ with respect to kinetic inductance $L_\text{k}$. The self-Kerr is extracted via global fit af different powers according to Appendix~\ref{app:eq}. Triangles [squares] represent resonators with inductors width of 500~nm [250~nm].
    In both panels, error bars represent the distribution of the $Q_i$ and $K$ fitted according to Eq.~\eqref{eq:s21_kerr_hanged} for fourteen devices [see also Fig.~\ref{fig:Qi_v_K_App}].}
    \label{fig:Lk_v_Qi_K}
\end{figure*}

Figure~\ref{fig:Lk_v_Qi_K}(a) reports the extracted quality factors of the 105 tested resonators in the linear regime (for Kerr-induced frequency shift much smaller than the resonator line width). The datapoints represent the averaged internal quality factor estimated for the seven resonators with different couplings $\kappa$ to the feedline [see Fig.~\ref{fig:Qi_v_K_App}(a) for the plot of the extracted $Q_i$ for all the measured resonators].
The error bars represent the standard deviation between the internal quality factor of the different devices. 
We also include in the error bar the uncertainty deriving from the global fit of each resonator, although we notice that they are negligible with respect to the parameter spreading due to fabrication.

To quantify the self-Kerr nonlinearity $K$ we reach regimes of large photon numbers, where the effect of the nonlinearity is comparable to the resonator linewidth.
In Fig.~\ref{fig:Lk_v_Qi_K}(b), we reported the measured $K$ of the critically coupled tested resonators ($\kappa \approx \gamma$) for each film and width. The data for the resonators with 500~nm (250~nm) wide inductor are shown with triangles (circles).
The self-Kerr nonlinearity increases with $L_\text{k}$, with a clear offset between the two inductor widths, as also reported in \cite{Anferov_2020}. 
This is in agreement with the relation
\begin{equation}
K \propto \frac{L_\text{k,0}~\omega_0^2}{(j_c wt)^{n_\text{fr}}}, \label{eq:chi_def}
\end{equation}
obtained by replacing $I_c = j_c wt$, in the Kerr equation [see Appendix~\ref{app:eq}, Eq.~\eqref{eq:chi_dep}], where $j_c$ is the critical current density of the thin film, and $w$ and $t$ are respectively the width and thickness of the inductor wire of the resonators. From this equation, we notice that the difference in width of the nanowire inductors clearly affects the value of $K$.

To highlight the dependencies of the self-Kerr $K$ from both kinetic inductance and inductor width, we plot in Fig.~\ref{fig:Qi_v_K_App}(b) $|K|/\omega_0^2$.

\subsection{Origin of the internal dissipation}

All devices, except the ones with the highest $L_\text{k}$, show $Q_i$ around $10^{5}$ in the single-photon regime. For the devices with the largest $L_\text{k} = 170$~pH/$\square$, we argue that the nitrogen concentration of the film is approaching the limiting value of the superconductor--normal metal transition (SNT) \cite{burdastyh_2022}. 
This argument is corroborated by the observation that the resonators with 250~nm wide inductors did not exhibit any superconducting transition while reaching base temperature, while the $w$~=~500~nm devices show a lower internal quality factor in the range of $10^{4}$ at single photon regime. 
These 170~pH/$\square$ resonators, however, present $Q_i$ in excess of $10^{5}$ at a thousand photon number, in line with the other films, suggesting that the dissipation can be mainly attributed to two-level system (TLS) hosted in highly disordered films \cite{McRae_2020, muller_2019}. While TLS nature is still unclear, it is believed to be related to atoms tunneling between two sites of a disordered solid \cite{Phillips_1972,Anderson_1972,Gao_thesis_2008}. 

\begin{figure*}
    \centering
    \captionsetup{justification=raggedright, singlelinecheck=false}
    \includegraphics[width=.96\linewidth]{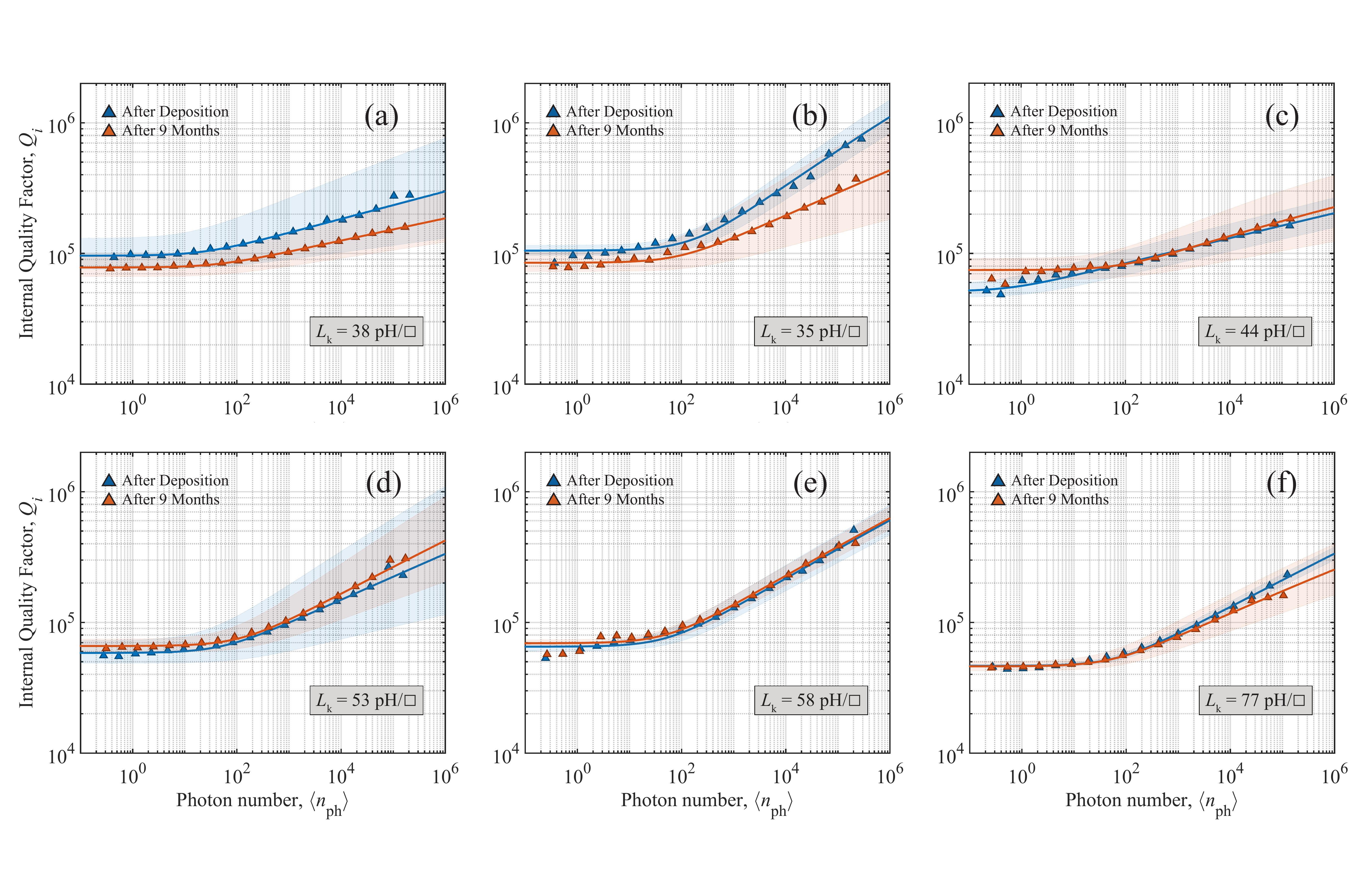} \vspace{-15pt}
    \caption{Internal quality factors $Q_i$ as a function of photon number for NbN compact resonators realized on different films. We performed power scans of critically coupled resonators of the several film compositions to estimate the TLS fitting parameters according to Eq.~\eqref{eq:qi_final}. Blue [red] datapoints and fitting functions represent $Q_i$ measured right after [nine months after] film deposition. The results suggest a clear dominance of TLS as main dissipation mechanism, as well as a more pronounced ageing effect for NbN films with lower N$_2$ concentration \textbf{(a,b)}.}
    \label{fig:ageing}
\end{figure*}

In order to describe the various mechanisms contributing to the internal quality factor as a function of input power, we write \cite{Pappas_2011,niepce_2019,scigliuzzo_2020}:
\begin{align}
    \frac{1}{Q_i} = &~\delta_0 + F\delta^0_\text{TLS} \frac{\tanh(\hbar\omega_0/2k_BT) }{(1+\braket{n_\text{ph}}/n_c)^\beta} + \nonumber
    \\ & + \frac{\alpha}{\pi} \sqrt{\frac{2 \Delta}{h f_r}} \frac{n_\text{qp}(T)}{n_s(0) \Delta}. \label{eq:qi_final}
\end{align}
In this equation, $\delta_0$ is the residual loss rate of the resonator, \textit{i.e.} the sum of the other loss contributions which are not described by TLS nor quasi-particle loss models. 
The second contribution is due to TLS, which according to the TLS model \cite{Gao_thesis_2008} generates a power and temperature dependent resonator loss. Here, $F$ is defined as the filling factor (the ratio between the electric field threading the TLS and the total electric field), $n_c$ is the characteristic photon number of TLS saturation, and $\delta^0_\text{TLS}$ is the intrinsic TLS loss. 
The third contribution is due to quasi-particles, where $\alpha$ is the ratio between kinetic and total inductance, $n_\text{qp}(T) = n_s(0) \sqrt{2 \pi k_B T \Delta} e^{-\Delta/k_B T}$ is the temperature dependent population of quasi-particles \cite{barends_2011}, and $n_s(0)$ is the Cooper pairs zero energy density of states.
We assumed $\alpha$ to be 1, as the kinetic inductance contribution to the resonant frequency largely dominates the geometric inductance.

Fig.~\ref{fig:ageing} reports the measured quality factor for representative resonators of different films, close to critical coupling conditions, $\kappa \approx \gamma$, as a function of the photon number. We then fit the model in Eq.~\eqref{eq:qi_final}, and we find that the TLS contribution dominates the internal quality factor. We report the value of $F\delta^0_\text{TLS}$ and $n_c$ obtained from the fits in Tab.~\ref{tab:devices}. Comparing them to what previously found in \cite{niepce_2019,Yu_2021}, we find similar fitting values for the exponent $\beta$, close to 0.2 for all the tested devices. However, as reported in Tab.~\ref{tab:devices}, the characteristic photon number of TLS saturation, $n_c$, is estimated to range between 30 and 100, a much larger value than the one reported for similar films in NbN \cite{niepce_2019}, which we attribute to a discrepancy in the photon number expressions [see Eq.~\eqref{eq:n}, in accordance with \cite{Yu_2021}].

\begin{figure*}
    \centering
	\captionsetup{justification=raggedright, singlelinecheck=false}
	\includegraphics[width=.96\linewidth]{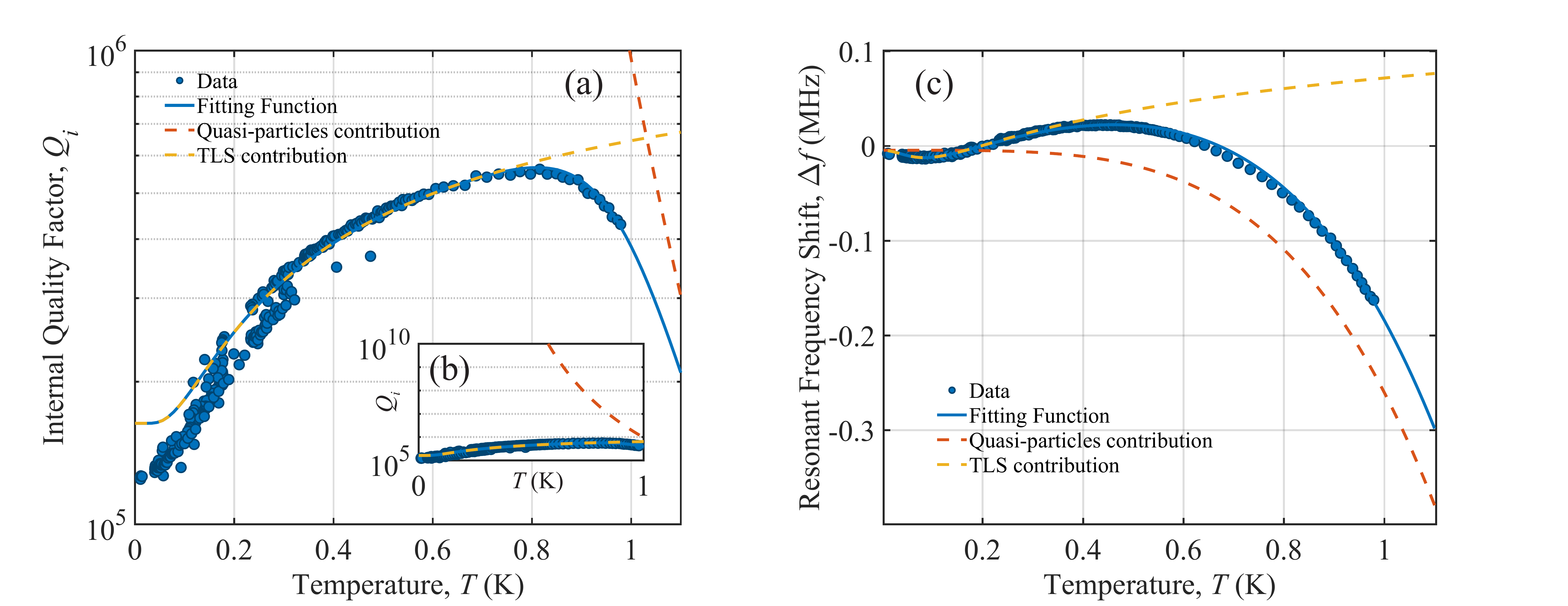}
	\caption{\textbf{(a)} Temperature dependence of internal quality factor $Q_i$ for device 803-500 [see Tab.~\ref{tab:devices}]. \textbf{(b)} Zoom--out of the panel \textbf{(a)} for $Q_i$ between $10^5$ and $10^{10}$, reported to highlight the expected evolution of quasi-particles contribution to losses. \textbf{(c)} Resonant frequency shift $\Delta f$ extracted from the same resonator analyzed in panel \textbf{(a)}. The validity of the quasi-particles and TLS parameters extracted from Eq.~\eqref{eq:qi_final} are confirmed by agreement with the resonant frequency shift $\Delta f$ predicted by Eq.~\eqref{eq:df_final}. Dashed red [yellow] lines represent the model of the contribution of quasi-particles [TLS] losses according to Eqs.~\eqref{eq:qi_final}--\eqref{eq:df_final} respectively in panels \textbf{(a)} and \textbf{(b)}.}
	\label{fig:Qi_v_T}
\end{figure*}

To further show the dominance of TLS, we investigated the resonator's quality factor evolution at different operating temperatures. The cryostat was slowly warmed up from a base temperature of 15~mK to a temperature of almost 1~K in a controlled way, and the resonator spectrum was acquired at $\braket{n_\text{ph}} \approx 50$. The evolution of the extracted internal quality factor $Q_i$ and resonator resonant frequency $\omega_0$ are reported in Fig.~\ref{fig:Qi_v_T}(a).

In addition, we characterize the contribution of the TLS and quasiparticle using the frequency shift of the resonators at different temperatures [see Fig.~\ref{fig:Qi_v_T}(b)]. Indeed, we have that \cite{scigliuzzo_2020}:
\begin{align}
    \frac{\Delta f}{f_r} = &~\frac{F\delta^0_\text{TLS}}{\pi} \bigg( \text{Re} \bigg\{ \Psi \bigg( \frac{1}{2}+\frac{h f_r}{2 i \pi k_B T} \bigg) \bigg\} + \nonumber 
    \\ & - \ln{\frac{h f_r}{2 \pi k_B T}} \bigg) - \alpha \frac{\Delta L_\text{k}}{L_\text{k}}, \label{eq:df_final}
\end{align}
where $\Psi$ is the digamma function \cite{Pappas_2011} and $\frac{\Delta L_\text{k}}{L_\text{k}}$ is the kinetic inductance change. Since $L_\text{k}(T) = \mu_0 \lambda^2(T) (l/wd)$, where $\lambda(T)$ is the temperature dependent London penetration depth, and as reported in \cite{lee_1993} for $T \le 0.4~T_C$, $[\lambda(T)/\lambda(0) - 1] \propto (T/T_C)^2$, we obtain
\begin{equation}
    \frac{\Delta L_\text{k}}{L_\text{k}} = \frac{\lambda^2 (T) - \lambda^2 (0)}{\lambda^2 (0)} \propto \bigg( \frac{k_B T}{\Delta} \bigg)^4.
\end{equation}

From the two fits of Eqs.~\eqref{eq:qi_final}--\eqref{eq:df_final} reported in Fig.~\ref{fig:Qi_v_T}, we extracted a thin film critical temperature $T_C$ of 7.4~K, and TLS contribution $F\delta^0_\text{TLS}$ of 1.8~$10^{-5}$, both in accordance with the previously obtained results reported in Fig.~\ref{fig:films} and Fig.~\ref{fig:ageing}.
We notice that for higher temperature the internal quality factor increases due to the saturation of TLS fluctuators, and then it starts dropping due to the losses caused by quasi-particles population. Due to the large $T_C$ of the characterized film ($T_C \approx 7.5~$K), this effect becomes dominant only at a temperature higher than 700~mK, \textit{i.e.} at roughly 10\% of $T_C$. The role played by TLS and quasi--particles is also confirmed by the temperature evolution of the resonator frequency shift $\Delta f$ [see Fig.~\ref{fig:Qi_v_T}(c)]. Below 400~mK $\Delta f$ is caused solely by TLS fluctuators, while at larger temperatures, the quasi-particles induced shift dominates.

\subsection{Ageing characterization}

Finally, to address the problem of thin films ageing, in particular the effect of the niobium oxide native layer on the internal quality factor \cite{santavicca_2015, medeiros_2019, verjauw_2021}, the devices were tested after nine months from the initial measurements, and in the same exact configuration. The devices were kept wire-bonded to the PCBs and were stored in a controlled N$_\text{2}$ atmosphere. We observed a systematic dip frequency shift of about $\Delta f_\text{age}$ ranging between 3 and 5 MHz [see Tab.~\ref{tab:devices}] towards lower frequency for all the measured resonators. In Fig.~\ref{fig:ageing} are collected the power scans of the internal quality factors of the tested resonators, with respect to average number of photons, measured right after fabrication [in blue] and after nine months [in red].
While for the low-$L_\text{k}$ devices the quality factor dropped slightly due to ageing, for larger $L_\text{k}$ films this effect is less pronounced, with the internal quality factor remaining almost unchanged after nine months. All the resonators remain, however, dominated by coupling to TLS, with internal quality factors at large number of photons in the excess of $10^{5}$. 

\section{\label{sec:conc}Discussion and Conclusion}

\begin{figure*}
    \centering
	\captionsetup{justification=raggedright, singlelinecheck=false}
	\includegraphics[width=.96\linewidth]{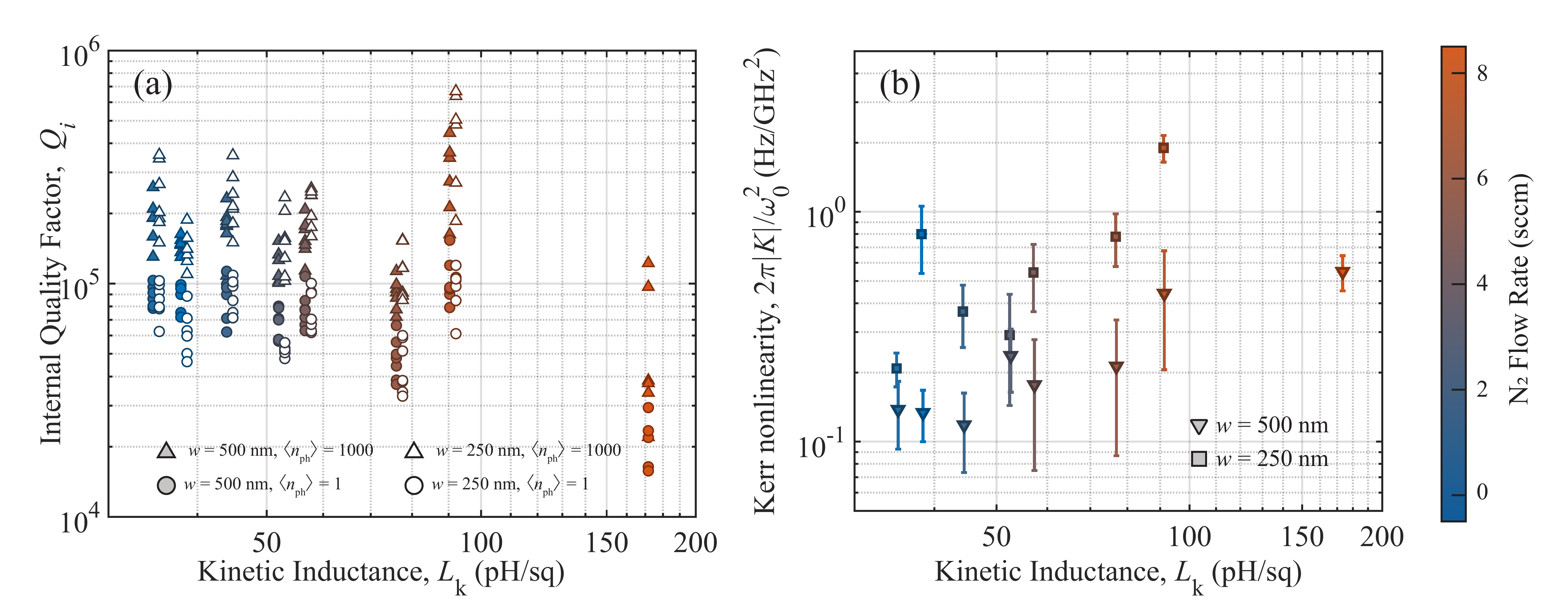}
	\caption{\textbf{(a)} Internal quality factor of all the 105 tested resonators with respect to kinetic inductance $L_\text{k}$, resonators widths and photon number, as reported in the legend. The horizontal drift between the full and empty datapoints is an artifact to simplify graph reading. \textbf{(b)} Plot of self-Kerr nonlinearity $K$ normalized with respect to $\omega_0^2$ to highlight the qualitative linear dependence on kinetic inductance.}
	\label{fig:Qi_v_K_App}
\end{figure*}

We have demonstrated high ($Q_i > 10^{5}$) internal quality factor, high--kinetic inductance superconducting resonators operating at single-photon regime based on NbN thin films superconductor. The NbN was bias-sputtered to increase kinetic inductance and device yield due to its poly-crystalline nature, at the expenses of a reduced critical temperature. 

Both stochiometric and non-stochiometric NbN films were deposited by varying atmospheric deposition conditions. After characterizing the deposition rates, films of equal thickness of $\sim$13~nm were sputtered. We fabricated and characterized 105 compact LC resonators multiplexed in hanger configuration and with kinetic inductance ranging from 30~pH/$\square$ to 170~pH/$\square$, as shown in Fig.~\ref{fig:Qi_v_K_App}(a). 
While for low photon number the internal quality factor presents a slight dependence from kinetic inductance, at high photon numbers the resonators quality factors approach $10^6$, suggesting a clear TLS induced loss mechanism.

By fitting the high--power scattering parameter of the resonators we also estimate the self-Kerr nonlinearity $K$ for the different tested films. Qualitatively, as depicted in Fig.~\ref{fig:Qi_v_K_App}(b), we observe a linear dependence of $K$ with respect to $L_\text{k}$. The sel-Kerr $K$, being below 100~Hz, is about four orders of magnitude lower than that of resonators made with Josephson junctions arrays \cite{krupko_2018}.

In conclusion, the results show a clear dominance of TLS--induced losses across the board, with lower kinetic--inductance films having larger internal quality factors, lower self-Kerr nonlinearity and less robustness with respect to ageing effects. The overall performance make the technology particularly appealing for those applications relying on low--nonlinearity, high--quality factor and large average photon number, such as parametric amplifiers, readout resonators, and photon detectors.

\section*{Contributions}

S.F. conceived the experiments with inputs from P.S., developed the recipes and optimized the deposition and characterization techniques. S.F., I.N.A. and F.O. fabricated the devices. S.F., S.Y.B.d.P. and V.J. measured the resonators. S.F. analyzed the data with inputs from M.S., F.M. and P.S.. S.F., F.M. and P.S. wrote the manuscript with inputs from all authors. E.C. supervised the work.

\section*{Acknowledgements}

This research was funded by Swiss National Science Foundation through the Sinergia programme grant number 177165, 2018—2022. P.S. acknowledges the project grants NCCR SNF 51AU40-1180604 and SNF project 200021-200418.

\appendix

\section{\label{app:lk}Estimation of kinetic inductance}

The kinetic inductance of the films was estimated both using classical BCS theory and via Sonnet simulations, by simulating the resonant frequency of the tested chips and comparing it to the measured value at low photon number. 

The chip designs were slightly different for the various films, to have the resonant frequencies of the devices in the same frequency range, and to ensure low reflections by matching as much as possible the microwave feedline to a 50~$\Omega$ impedance line. We used overall five designs.

\begin{figure*}
    \centering
	\captionsetup{justification=raggedright, singlelinecheck=false}
	\includegraphics[width=.96\linewidth]{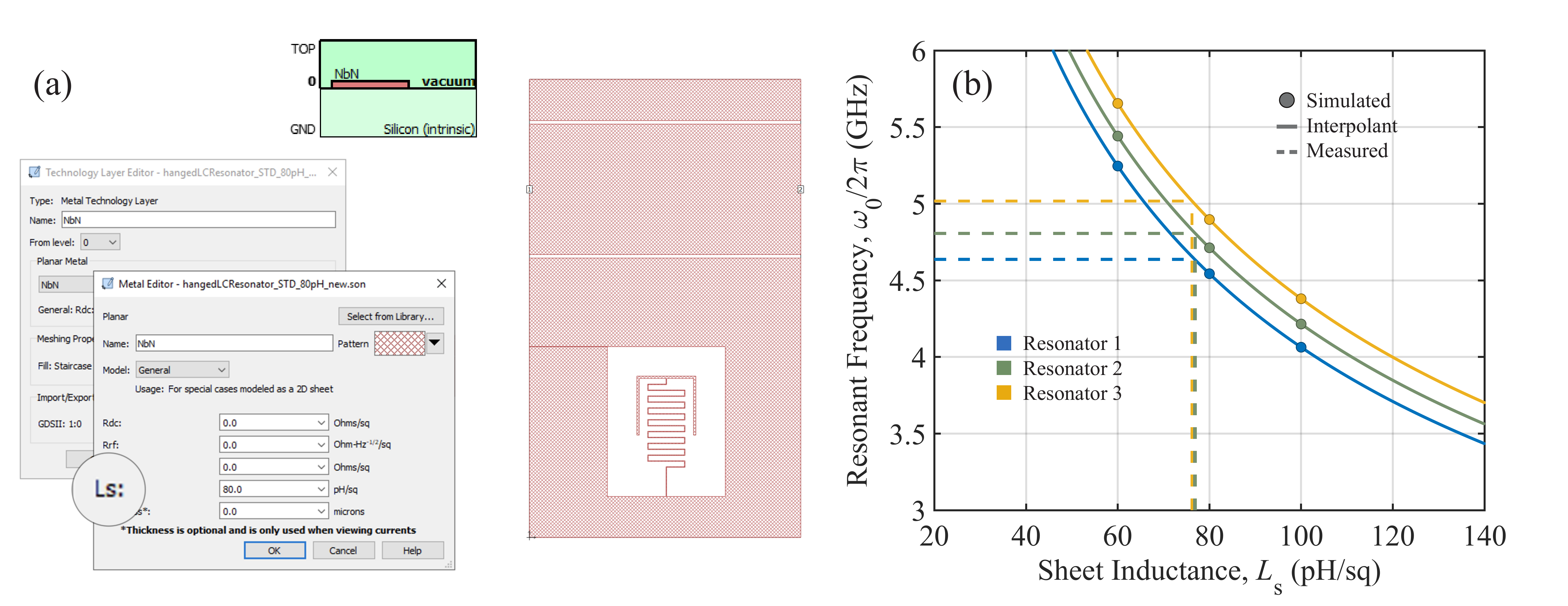}
	\caption{\textbf{(a)} Sonnet simulation setup. The NbN layer was modeled as a perfect conductor with variable sheet inductance $L_\text{s}$. \textbf{(b)} Interpolated kinetic inductivity -- resonant frequency curve for the three simulated resonators. Measuring the resonant frequency of each of the three simulated devices, it is possible to extract the kinetic inductivity of the superconducting film as the average of the three $L_\text{s}$ values.}
	\label{fig:Lk_vs_f_App}
\end{figure*}

For each design, we simulated three resonators with Sonnet to estimate their resonant frequency at three different sheet kinetic inductivity values $L_\text{s}$, see Fig.~\ref{fig:Lk_vs_f_App}(a). Given that the resonator capacitance remains constant, the resonant frequency should follow a $\omega_0 \propto 1/\sqrt{L_\text{k}}$ dependence. By plotting the hyperbola passing through these three points, as shown in Fig.~\ref{fig:Lk_vs_f_App}(b), it is possible, by measuring at low photon number the resonant frequency of the resonator under test $\omega_0/2\pi$, to estimate the kinetic inductance of the film. To increase the robustness of such method, we average the obtained $L_\text{k}$ for the three resonators [see Fig.~\ref{fig:Lk_vs_f_App}], allowing us to reduce even further the uncertainty associated to the kinetic inductivity of the films.

\section{\label{app:eq}Equations derivation}

The transmission spectra from the main feedline was measured at different input power. The nonlinear effects of the kinetic inductance of the film, produces an effective Kerr behavior of the resonators (see Fig.~\ref{fig:resonator}). 

The nonlinear Kerr Hamiltonian for superconducting resonators can be written as \cite{Anferov_2020}:
\begin{equation}
    H = \hbar \omega_0 a^\dagger a + \hbar \frac{K}{2} (a^\dagger a)^2, \label{eq:nonlin_hamilt}
\end{equation}
where $K \propto \Delta L \omega_0^2$ is Kerr--nonlinearity, $\Delta L$  inductance change and $\omega_0$  fundamental resonator frequency. Following the same formalism of \cite{Eichler_2014, Anferov_2020} using input--output theory leads to the scattering parameter for hanger resonators of:
\begin{equation}
    S_{21} = 1 - \frac{\kappa}{\kappa+\gamma}\frac{e^{i\phi}}{\cos{\phi}}\frac{1}{1+2i(\delta-\xi n)}, \label{eq:s21_kerr_hanged_app}
\end{equation}
with
\begin{equation}
    \delta \equiv \frac{\omega_\text{d}-\omega_0}{\kappa+\gamma}, \quad \xi \equiv \frac{|\tilde{\alpha}_{in}|^2 K}{\kappa+\gamma}, \quad n \equiv \frac{|\alpha|^2}{|\tilde{\alpha}_{in}|^2}, 
\end{equation}
$\tilde{\alpha}_{in} \equiv \sqrt{\kappa}\alpha_{in}/(\kappa+\gamma)$, $\kappa \equiv \omega_0 / Q_\text{c}$ being the coupling to the feedline, $\gamma \equiv \omega_0 / Q_\text{i}$ the internal resonator losses. $\phi$ is a parameter used to take into account impedance mismatch between the feedline and the resonator [also called the $\phi$ rotation method ($\phi$RM) \cite{Gao_thesis_2008}] and $n$ is calculated as solution of the equation:
\begin{equation}
    \frac{1}{2} = \bigg( \delta^2+\frac{1}{4}\bigg)n - 2\delta \xi n^2 + \xi^2 n^3. \label{eq:number_photons}
\end{equation}

Cryogenic attenuated input lines were calibrated at base temperature and an extra contribution of -8~dB was estimated from the coaxial lines, for a total attenuation of -68~dB. Nonlinear effects described hereon have been analyzed from the assumption that the input power is known within an error of $\pm$2~dB. 
The average photon number $\braket{n_\text{ph}} = |\tilde{\alpha}_{in}|^2$ in the resonator 
\begin{equation}
    |\tilde{\alpha}_{in}|^2 = \frac{\kappa}{(\kappa + \gamma)^2}\frac{P_\text{in}}{\hbar\omega_0}, \label{eq:n}
\end{equation}
together with Eq.~\eqref{eq:n}, allows to estimate the Kerr nonlinearity $K$ from the fitting functions parameters $\xi$ and $n$:
\begin{equation}
    K = \xi \frac{(\kappa+\gamma)}{|\tilde{\alpha}_{in}|^2}. \label{eq:chi}
\end{equation}
Using the generalized relation of current dependence of kinetic inductance derived in \cite{clem_kinetic_2012}, in the limit for small $I/I_c$, we get:
\begin{align}
    L_\text{k} &~= L_\text{k,0}~(1-(I/I_c)^{n_\text{fr}})^{-1/{n_\text{fr}}} \nonumber
    \\ &~\approx L_\text{k,0}~\bigg(1 + \frac{1}{n_\text{fr}}\bigg(\frac{I}{I_c}\bigg)^{n_\text{fr}}\bigg). \label{eq:lk_dep}
\end{align}
Substituting Eq.~\eqref{eq:lk_dep} in the relation for Kerr:
\begin{equation}
    K \propto \Delta L~\omega_0^2 \propto \frac{L_\text{k,0}~\omega_0^2}{I_c^{n_\text{fr}}}, \label{eq:chi_dep}
\end{equation}
where $L_\text{k,0}$ is the kinetic inductance of the superconducting film at zero bias current, $n_\text{fr} = 2.21$ is the exponential of the fast relaxation limit function defined in \cite{clem_kinetic_2012}, $I$ is the bias current, which is proportional to the pump power $P_\text{in}$.
In order to compare the nonlinearity of the tested devices, we have measured the critical current of the coupling feedline so to have an estimate of the critical current $I_c$. 

\begin{table*}
    \centering
	\captionsetup{justification=raggedright, singlelinecheck=false}
	\caption{Superconducting devices main parameters for the films presented in Tab.~\ref{tab:films}. For each film composition and resonator inductor width, the table reports only the values of the devices closest to critically coupled conditions ($\kappa \approx \gamma$).}
	\begin{tabularx}{\textwidth}{XYYYYYYYY}
		\hline
		\multicolumn{2}{>{\hsize=\dimexpr2\hsize+2\tabcolsep+\arrayrulewidth\relax}Y}{\textbf{Device Design}} &  \multicolumn{7}{>{\hsize=\dimexpr7\hsize+7\tabcolsep+\arrayrulewidth\relax}Y}{\textbf{Device Properties}} \\
	    & Width & $\omega_0 / 2 \pi$ & $\kappa / 2 \pi$ & $\gamma / 2 \pi$ & $K / 2 \pi$ & $F\delta^0_\text{TLS} $ & $n_c$ & $\Delta f_\text{age} $ \\ [-3pt]
		ID & (nm) & (GHz) & (kHz) & (kHz) & (Hz) & $(\times10^{-5})$ & & (MHz) \\
		[.5\normalbaselineskip] \hline
		801-500 & 500 & 5.58 & 86.95 & 57.52 & - 2.40 & 1.02 $\pm$ 0.16 & 34.14 $\pm$ 12.97 & 3.42 \\
		\rowcolor{gray!10}801-250 & 250 & 4.95 & 112.87 & 68.00 & - 19.58 & 1.40 $\pm$ 0.05 & 9.69 $\pm$ 4.35 & --\\
		802-500 & 500 & 5.87 & 74.51 & 61.77 & - 3.05 & 1.00 $\pm$ 0.03 & 47.07 $\pm$ 11.08 & 3.78 \\
		\rowcolor{gray!10}802-250 & 250 & 5.17 & 40.59 & 66.21 & - 5.56 & 1.22 $\pm$ 0.27 & 42.84 $\pm$ 48.82 & --\\
		803-500 & 500 & 5.95 & 91.88 & 95.96 & - 4.17 & 1.83 $\pm$ 0.12 & 1.74 $\pm$ 1.58 & 4.83 \\
		\rowcolor{gray!10}803-250 & 250 & 5.15 & 50.42 & 66.66 & - 11.50 & 1.31 $\pm$ 0.07 & 44.38 $\pm$ 16.56 & --\\
		804-500 & 500 & 5.49 & 123.21 & 95.31 & - 7.13 & 1.69 $\pm$ 0.20 & 77.03 $\pm$ 34.87 & 4.00 \\
		\rowcolor{gray!10}804-250 & 250 & 4.74 & 45.52 & 90.26 & - 9.59 & 1.88 $\pm$ 0.094 & 64.42 $\pm$ 22.21 & --\\
		805-500 & 500 & 5.72 & 105.78 & 91.09 & - 5.77 & 1.53 $\pm$ 0.15 & 47.82 $\pm$ 29.05 & 3.76 \\
		\rowcolor{gray!10}805-250 & 250 & 5.06 & 59.69 & 73.91 & - 13.94 & 1.41 $\pm$ 0.07 & 51.64 $\pm$ 14.33 & --\\
		806-500 & 500 & 4.93 & 76.049 & 111.30 & - 5.17 & 2.16 $\pm$ 0.17 & 48.56 $\pm$ 18.93 & 3.28 \\
		\rowcolor{gray!10}806-250 & 250 & 4.37 & 64.83 & 132.39 & - 14.88 & 2.84 $\pm$ 0.19 & 22.71 $\pm$ 12.56 & --\\
		807-500 & 500 & 6.47 & 64.81 & 42.17 & - 18.46 & 0.67 $\pm$ 0.05 & 94.46 $\pm$ 50.58 & --\\
		\rowcolor{gray!10}807-250 & 250 & 6.59 & 49.06 & 62.92 & - 82.39 & 1.43 $\pm$ 0.63 & 2.23 $\pm$ 4.27 & --\\
		808-500 & 500 & 3.34 & 45.63 & 233.55 & - 6.23 & 6.72 $\pm$ 0.29 & 30.84 $\pm$ 12.42 & -- \\ \hline
	\end{tabularx} 
	\label{tab:devices}
\end{table*}

\bibliographystyle{aipnum4-1}
\bibliography{Bibliography_Cryogenics.bib,Bibliography_SNSPD.bib,Bibliography_SuperconductingCircuits.bib,Bibliography_SCresonators.bib,Bibliography.bib}

\end{document}